\documentclass[letterpaper,ngerman,english]{article}
\usepackage[T1]{fontenc}
\usepackage[latin9]{inputenc}
\usepackage{float}
\usepackage{textcomp}
\usepackage{graphicx}

\makeatletter

\providecommand{\tabularnewline}{\\}

\newcommand{\lyxaddress}[1]{
	\par {\raggedright #1
	\vspace{1.4em}
	\noindent\par}
}

\makeatother

\usepackage{babel}
\begin{document}
\date{}
\title{Fusion of time of arrival and time difference of arrival for ultra-wideband
indoor localization }
\author{Juri Sidorenko \textsuperscript{ab*}, Volker Schatz\textsuperscript{a},\\
Norbert Scherer-Negenborn \textsuperscript{a}, Michael Arens \textsuperscript{a},
\\
Urs Hugentobler \textsuperscript{b}}
\maketitle

\lyxaddress{\textsuperscript{a}Fraunhofer Institute of Optronics, System Technologies
and Image Exploitation IOSB, Germany - juri.sidorenko@iosb.fraunhofer.de\\
\textsuperscript{b}Institute of Astronomical and Physical Geodesy,
Technical University of Munich, Germany - urs.hugentobler@tum.de}

Keywords: time of arrival, time difference of arrival, localization,
positioning, navigation, two-way ranging, Decawave, TWR, TOA, TDOA

\section{Abstract}

This article presents a new approach for the wireless clock synchronization
of Decawave ultra-wideband transceivers based on the time difference
of arrival. The presented techniques combine the time-of-arrival and
time-difference-of-arrival measurements without losing the advantages
of each approach. The precision and accuracy of the distances measured
by the Decawave devices depends on three effects: signal power, clock
drift, and uncertainty in the hardware delay. This article shows how
all three effects may be compensated with both measurement techniques. 

\section{Introduction}

Localization systems have become indispensable for everyday life.
Satellite navigation\cite{GPS1,GPS2} has displaced paper maps and
is now essential for the autonomous operation of cars and airplanes.
As the requirements of logistics and manufacturing processes increase,
access to precise positional information is becoming a necessity.
Depending on the operating conditions for the localization application,
different measurement principles \cite{LPM,RSSI,UWB_Limits} and techniques
\cite{TOA,TDOA,angleOfarrival} are available. Two of the most common
measurement techniques are based on the time of arrival (TOA) \cite{TOA}
and the time difference of arrival (TDOA) \cite{TDOA}. TOA calculates
the distance between two stations from the signal traveling time,
whereas TDOA considers the travel time differences between the stations.
Two-way ranging (TWR) uses TOA to calculate the distance between two
stations \cite{TWR}. In contrast to the one-way ranging approach
used by satellite-based applications, the TWR approach includes a
response to the transmitted signal. As a result, the transmitting
stations are not required to be synchronous. In applications where
not just the distance but also the position of the target (Tag) with
respect to the other stations (Anchors) needs to be known, TWR is
less suitable due to its slow update rate. Trilateration in two-dimensional
space requires at least three distance measurements. As the number
of tags increases, the update rate decreases. In contrast to TOA,
TDOA remains suitable for applications with large numbers of tags.
In TDOA applications, the anchors do not respond the tags. Multilateration
is performed by considering time stamp differences between anchors.
Geometrically, TOA equations describe circles, whereas TDOA equations
are hyperbolas in a two-dimensional space. Much like satellite navigation
systems, which are based on TOA, the clocks of the TDOA anchors must
be synchronized. This synchronization can be performed by wire \cite{TDOA_WIRE}
or with an additional station \cite{LPM}. \\
The measuring equipment is just as important as the measurement technique
itself. This article focuses on indoor radio frequency (RF)-based
localization systems. In general, indoor positioning applications
are a challenge for RF-based localization systems. Reflections can
generate interference with the main signal and lead to fading. Compared
to narrowband signals, ultra-wideband (UWB) signals are more robust
against fading \cite{fading1,fading2}. The Decawave transceiver \cite{Why_Decawave}
uses ultra-wideband (UWB) technology and is compliant with the IEEE802.15.4-2011
standard \cite{Decawave_Anaysis}. It supports six frequency bands
with center frequencies from 3.5 GHz to 6.5 GHz and data rates of
up to 6.8 Mb/s. Depending on the selected center frequency, the bandwidth
ranges from 500 to 1000 MHz. Various methods for wireless TDOA clock
synchronization are presented in \cite{TDOA_Sync1,TDOA_Sync2,TDOA_sync3}.
One aspect shared by all of them is that they use a fixed and known
time interval for the synchronization signal. In our case, the synchronization
signal is part of the localization and the time interval does not
need to be known. The solution presented here merges TOA and TDOA
measurements to increase the number of equations without losing the
specific advantages of each method. The measurements are provided
by Decawave EVK1000 transceivers without additional synchronization
hardware. This system can operate in indoor environments due to its
ability to deal with fading. The precision and accuracy of the Decawave
UWB depend primarily on three factors: the received signal power,
the clock drift, and the hardware delay. In {[}Journal: signal power
calibration{]}, we showed how the signal power correction curve can
be obtained automatically and how the clock drift can be corrected
in every measurement. In the present publication, we demonstrate how
to apply these corrections for TOA and TDOA localization.

\begin{table}[H]
\begin{centering}
\caption{Notations used\label{tab:Exmplation-of-the}}
\ \\
\par\end{centering}
\centering{}%
\begin{tabular}{|c|c|}
\hline 
Notations & \selectlanguage{ngerman}%
Definition\selectlanguage{english}%
\tabularnewline
\hline 
\hline 
$T_{i}^{R}$ & Time stamp at the reference station\tabularnewline
\hline 
$T_{i}^{T}$ & Time stamp at the tag\tabularnewline
\hline 
$T_{i}^{S}$ & Time stamp at the anchor station\tabularnewline
\hline 
$\Delta T_{n,m}$ & Difference between two time stamps $T_{m}-T_{n}$\tabularnewline
\hline 
$C_{n,m}$ & Clock drift error calculated from the time stamps n and m\tabularnewline
\hline 
$E_{i}$ & Time stamp error due to the signal power\tabularnewline
\hline 
$A,B$ & Hardware delay\tabularnewline
\hline 
$K$ & Time difference between the reference station and the tag \tabularnewline
\hline 
$x_{R},y_{R},z_{R}$ & Position of reference station \tabularnewline
\hline 
$x_{S},y_{S},z_{S}$ & Positions of base stations \tabularnewline
\hline 
$x_{T},y_{T},z_{T}$ & Position of the tag \tabularnewline
\hline 
$c_{0}$ & Speed of light\tabularnewline
\hline 
\end{tabular}
\end{table}

\section{Clock drift and signal power correction}

\begin{figure}[H]
\begin{centering}
\includegraphics[scale=0.4]{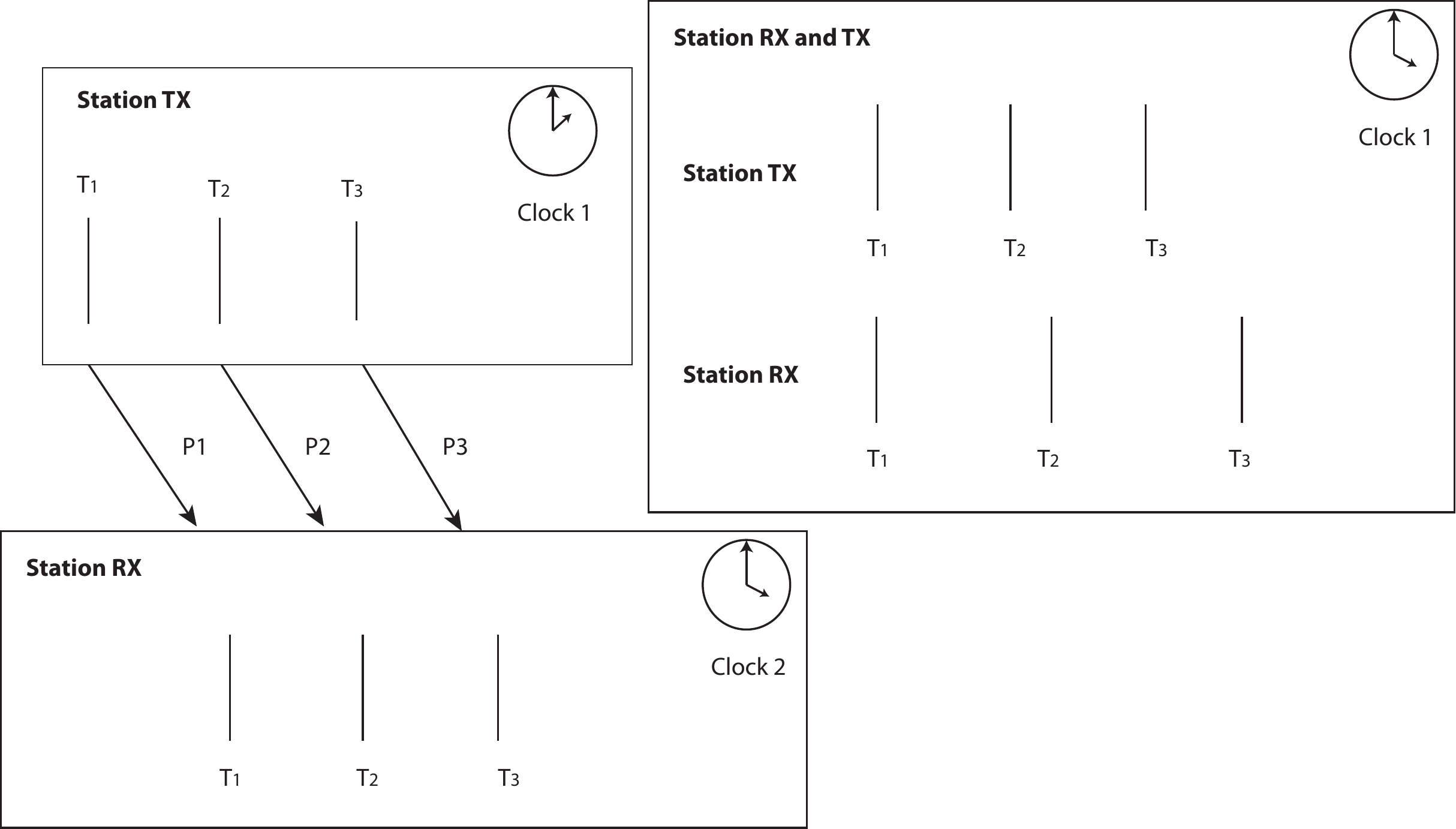}
\par\end{centering}
\caption{An alternative clock drift correction \label{fig:Alternative-clock-drift}{[}JOURNAL:
SIGNAL POWER{]}}
\end{figure}

In {[}Journal: signal power calibration{]}, we described how the clock
drift and the signal power correction can be determined for the Decawave
UWB transceivers. This section gives a short overview of the methods
used. Figure \ref{fig:Alternative-clock-drift} shows how the clock
drift can be corrected with linear interpolation. The transmitting
station (TX) sends three signals at times $T_{1}$,$T_{2}$ and $T_{3}$.
The clocks of the transmitter and the receiver are not synchronous.
If the clocks have no drift, then both clocks have the same frequency,
and the difference $\Delta T_{1,2}=T_{2}-T_{1}$ is the same for both
the transmitter and the receiver. If not, $\Delta T_{1,2}^{RX}\neq\Delta T_{1,2}^{TX}$
. The same principle applies to $\Delta T_{1,3}$. If the clock of
the reference station (RX) is running faster than the clock of the
transmitter station TX, then $\Delta T_{1,3}^{RX}>\Delta T_{1,3}^{TX}$
, and the clock drift error is equal to $C_{1,3}=\Delta T_{1,3}^{RX}-\Delta T_{1,3}^{TX}$
. Using linear interpolation, we can estimate the shift of the timestamp
$T_{2}$ due to clock drift. The correction term is equal to $\frac{C_{1,3}}{\Delta T_{1,3}^{TX}}\cdotp\Delta T_{1,2}^{TX}$
. A position error caused by a constant velocity of the object is
also corrected by the linear interpolation, due to the linear increase
of the position error (pseudo clock drift). In pratise, is $\Delta T_{1,3}^{TX}$
about 1 ms. An acceleration high enough to cause an error greater
than 5 mm, would require near most 1,000g $\left(10^{4}\frac{m}{s^{2}}\right)$.
The standard approach uses the integral of the phase-locked loop (PLL)
to calculate the correction value. In {[}Journal: signal power calibration{]},
we showed that this correction method may not be suitable due to its
dependency on the signal power. Alternative methods such as symmetric
and asymmetric double-sided two-way ranging \cite{TWR} do not calculate
the clock drift but use three or more messages to reduce the error. 

\begin{figure}[H]
\begin{centering}
\includegraphics[scale=0.4]{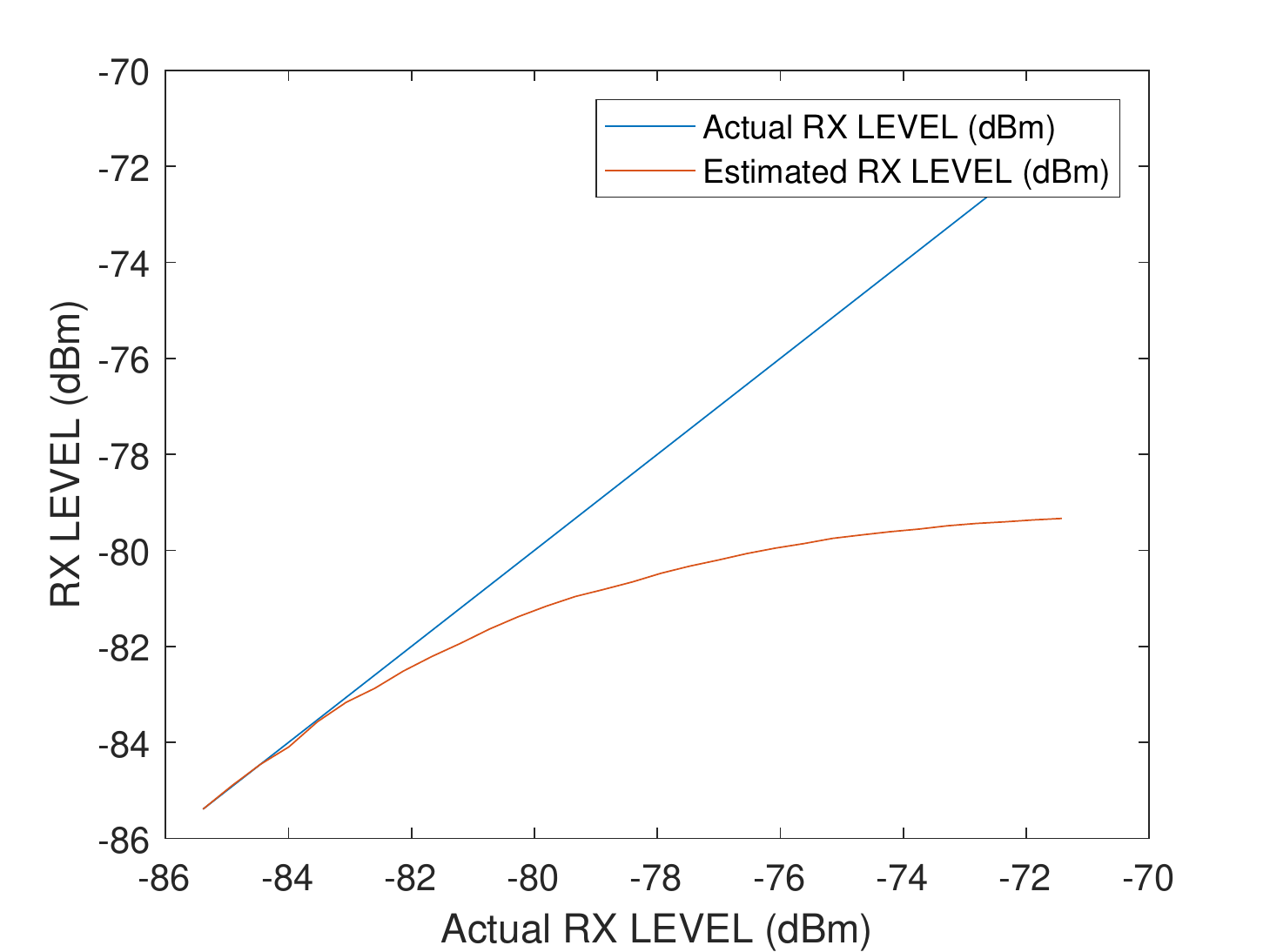}\includegraphics[scale=0.4]{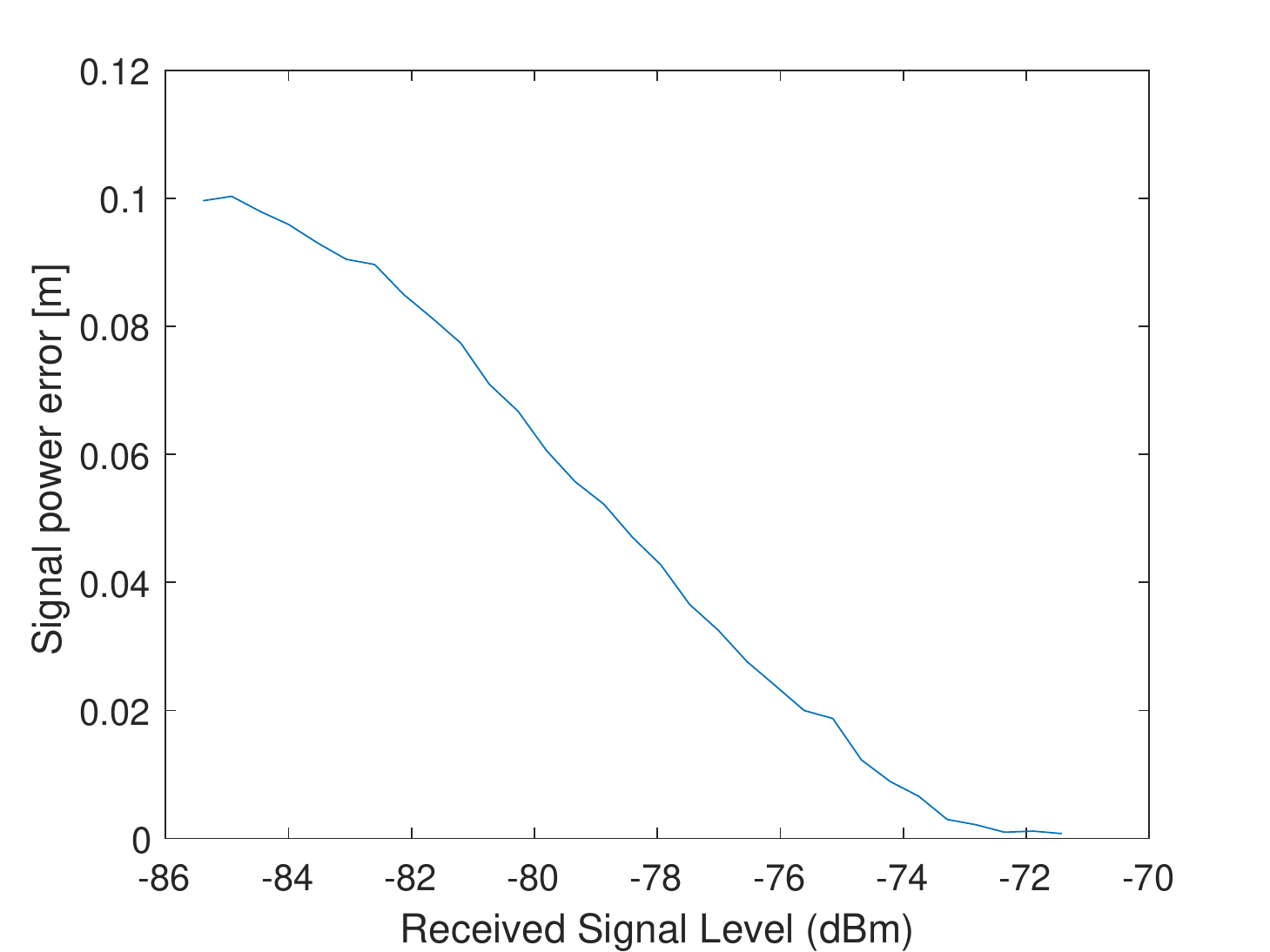}
\par\end{centering}
\caption{Final results of the power correction \label{fig:Final-results-of}.
Left: Measured signal power vs. real signal power. Right: Signal power
error curve {[}JOURNAL: SIGNAL POWER{]}}
\end{figure}

The timestamp of the DW1000 is known to be affected by the signal
power \cite{Signal_power,SignalPower_correction}. Increasing the
signal power causes a negative shift of the time stamp and vice versa.
In {[}Journal{]}, we showed how the signal power correction curve
can be determined for each Decawave UWB transceiver individually without
requiring additional measurement equipment. Figure \ref{fig:Final-results-of}
shows the correction curves for the measured vs. the actual signal
power and the actual signal vs. the timestamp error. 

\section{Time of arrival}

\begin{figure}[H]
\begin{centering}
\includegraphics[scale=0.4]{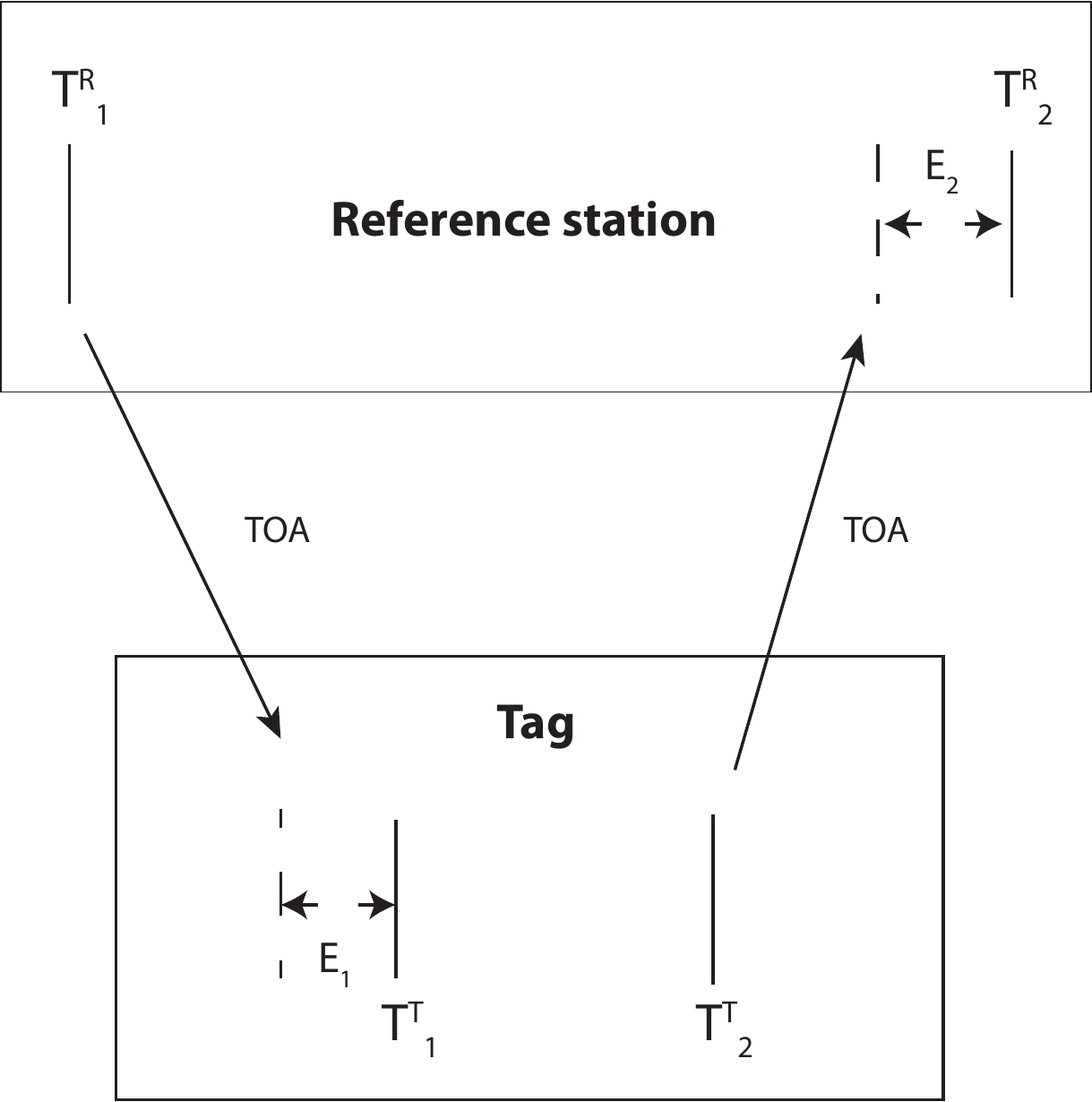}\ \ \ \ \ \ \includegraphics[scale=0.4]{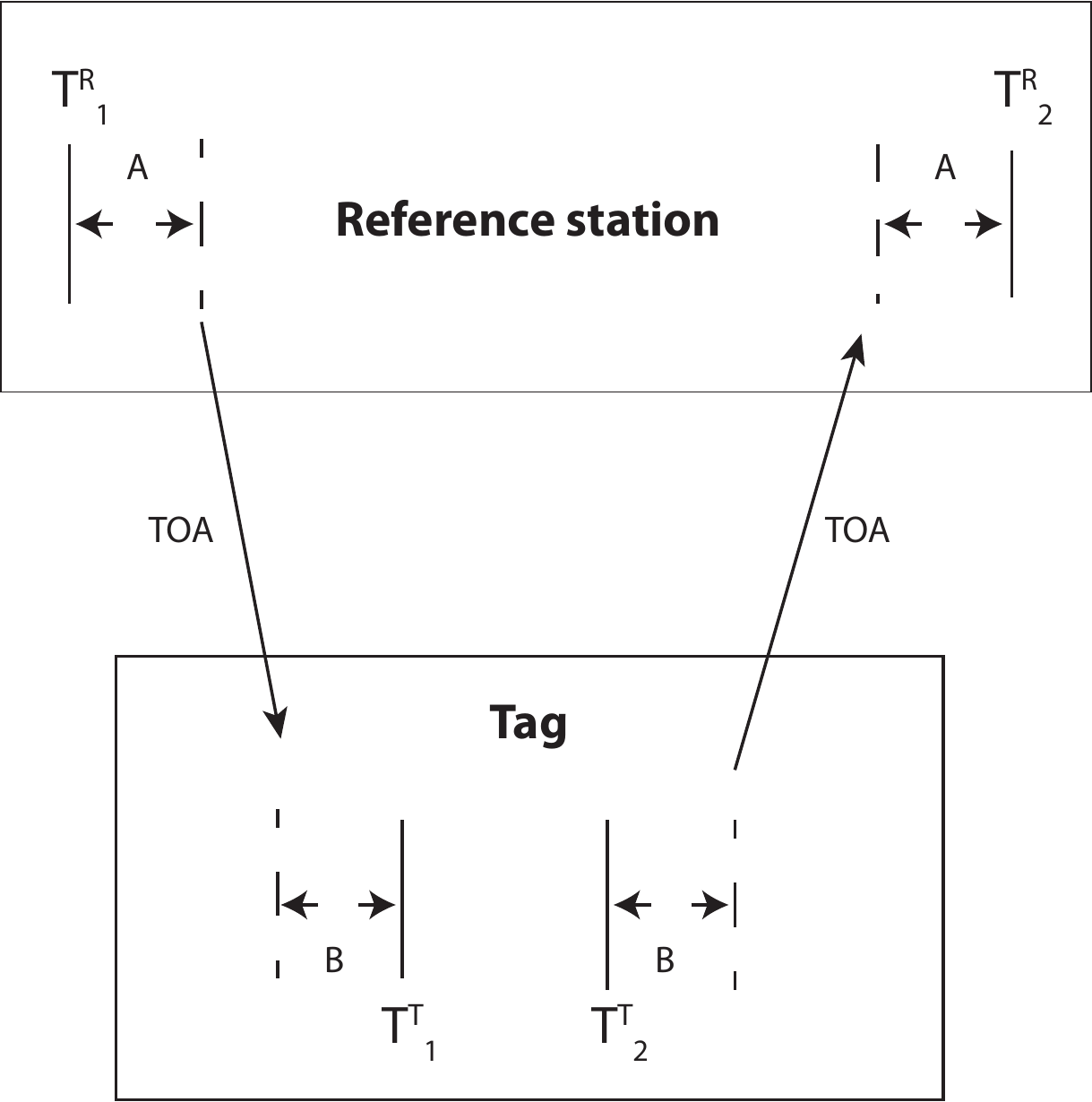}
\par\end{centering}
\caption{Left: Effect of the power on the TOA, Right: Effect of the hardware
delay on TOA\label{fig:TWR}}
\end{figure}

Figure \ref{fig:TWR} illustrates the concept of TWR and the timestamp
shift caused by signal power, as well as the error due to hardware
delay. In our implementation, the reference station is the initiator.
The first message is sent by the reference station with timestamp
$T_{1}^{R}$. The timestamp of the received message at the tag is
affected by the signal power, resulting in a timestamp shift of $E_{1}$.
The same applies to the response message, this time at the reference
station. It is important to note that the timestamps $T_{1}^{R}$
and $T_{2}^{T}$ are not affected by the receiving signal power. However,
the hardware delay (A,B) must always be considered. The sending delay
is assumed to be equal to the receiving delay. Without correction,
the TWR signal travel time is $0.5\cdotp\left(\left(T_{2}^{R}-T_{1}^{R}\right)-\left(T_{2}^{T}-T_{1}^{T}\right)\right)$. 

The corrected time of flight between the reference station and the
tag can be estimated with the following formula. 

\begin{equation}
T_{TOA}=0.5\cdotp\left(\left(T_{2}^{R}-T_{1}^{R}\right)-\left(T_{2}^{T}-T_{1}^{T}\right)-E_{2}-E_{1}\right)-A-B
\end{equation}

\begin{figure}[H]
\begin{centering}
\includegraphics[scale=0.45]{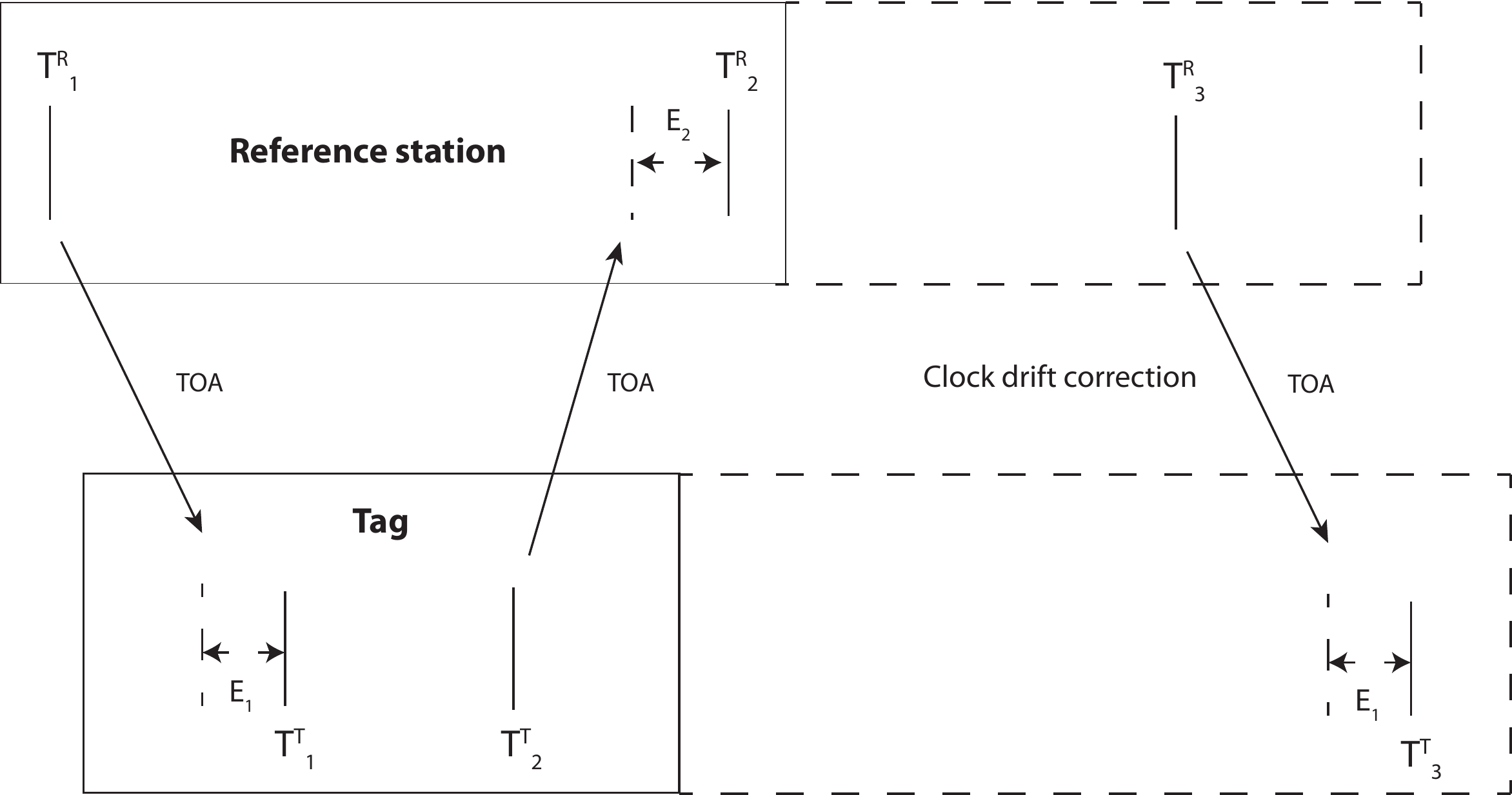}
\par\end{centering}
\caption{TWR clock drift correction \label{fig:TWR-clock-drift}}
\end{figure}

The values $E_{1}$ and $E_{2}$ are deduced from the signal power
correction curve. Note that the signal power may affect the tag and
the reference station differently. At lower signal power, the time
difference $\Delta T_{1,2}^{R}$ increases.\\
In the previous section, we showed that the clock drift can be corrected
by three messages. Figure \ref{fig:TWR-clock-drift} demonstrates
how this principle can be adapted for two-way ranging. The last message
is used to calculate the clock drift error $C_{1,3}^{RT}=\Delta T_{1,3}^{R}-\Delta T_{1,3}^{T}$.
Observe that the signal power $E_{1}$ does not affect the timestamp
difference $\Delta T_{1,3}^{T}$. The final time of flight equation
with the clock drift correction and three messages is as follows: 

\begin{equation}
T_{TOA}=0.5\cdotp\left(\Delta T_{1,2}^{R}-\Delta T_{1,2}^{T}-\left(\frac{C_{1,3}^{RT}}{\Delta T_{1,3}^{T}}\cdotp\left(\Delta T_{1,2}^{T}+E1\right)\right)-E_{2}-E_{1}\right)-A-B
\end{equation}

Given the corrected time measurements, the TOA, and the propagation
speed of the signal, lateration may be performed to deduce the position
of the tag $(x_{T},y_{T},z_{T})$ with respect to the anchors, by
solving the following system of equations. 
\begin{center}
\begin{tabular}{ccc}
$T_{TOA_{i}}\cdotp c_{0}=\sqrt{\left(x_{R_{i}}-x_{T}\right)^{2}+\left(y_{R_{i}}-y_{T}\right)^{2}+\left(z_{R_{i}}-z_{T}\right)^{2}}$ &  & $1\leq i\leq N$\tabularnewline
\end{tabular}
\par\end{center}

\section{Time difference of arrival}

\begin{figure}[H]
\begin{centering}
\includegraphics[scale=0.4]{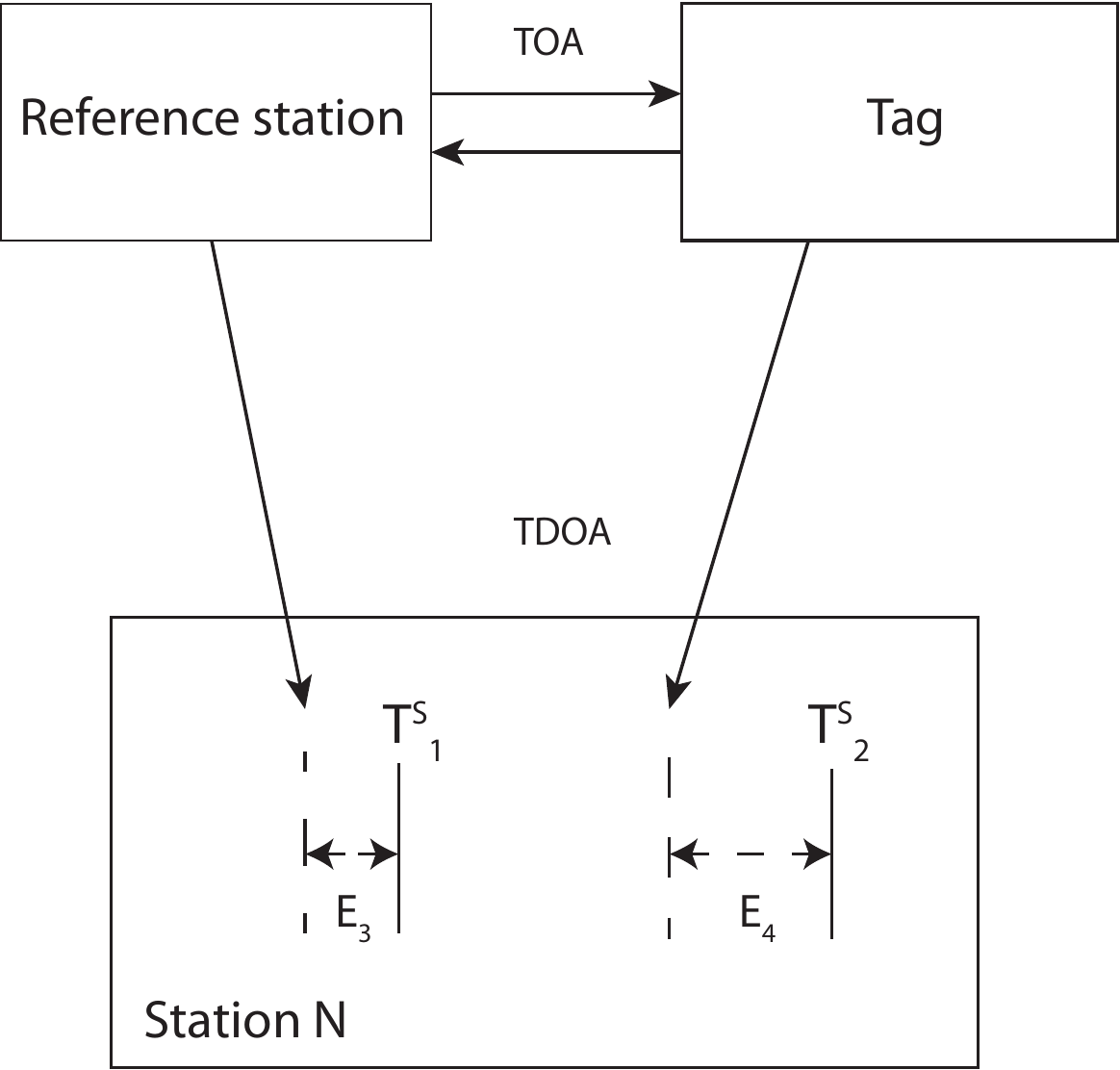}\ \ \ \ \ \ \includegraphics[scale=0.4]{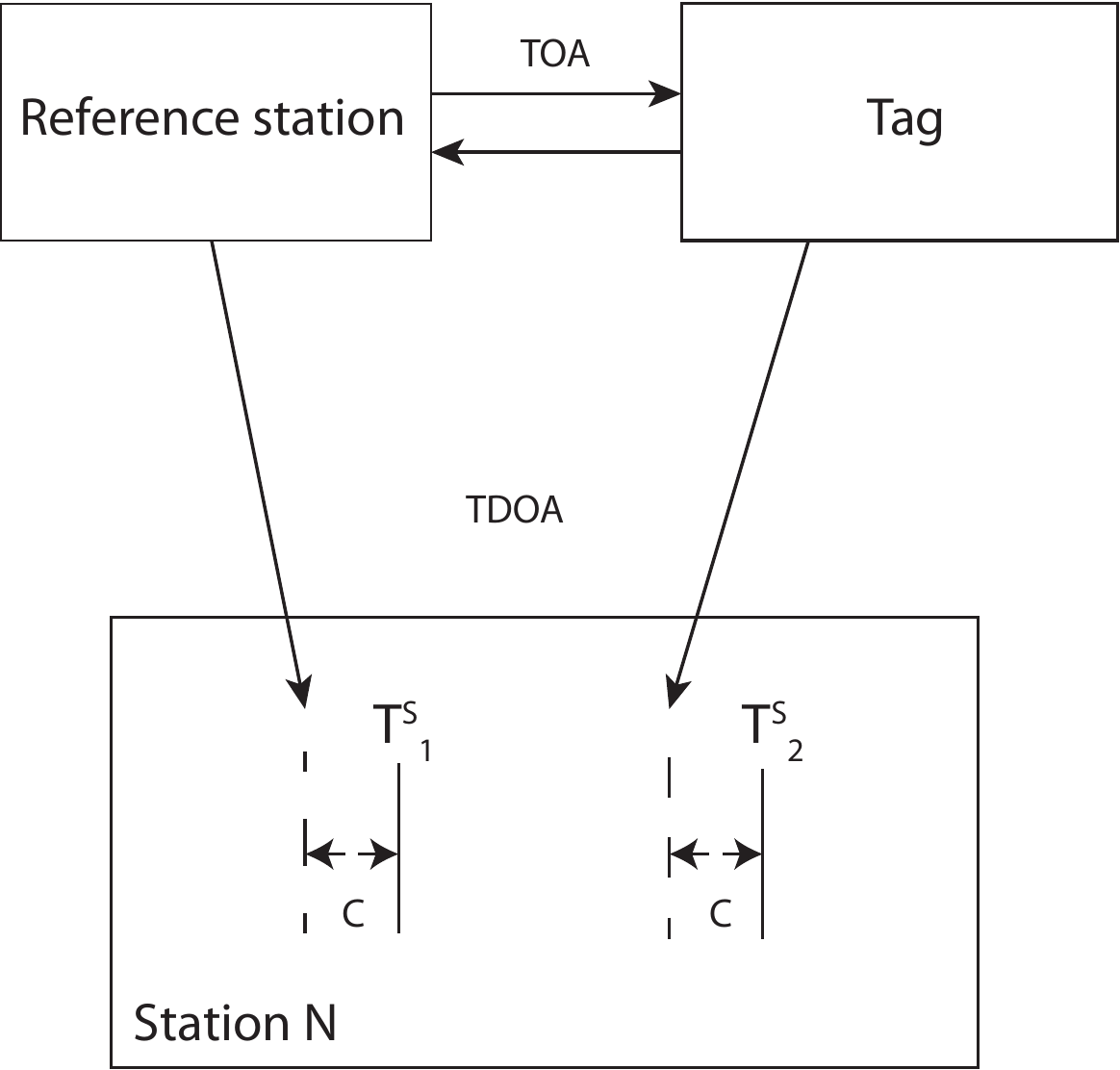}
\par\end{centering}
\caption{Left: Effect of power on the TDOA, Right: Effect of the hardware offset
on the TDOA \label{fig:TDOA}}
\end{figure}

\begin{figure}[H]
\begin{centering}
\includegraphics[scale=0.4]{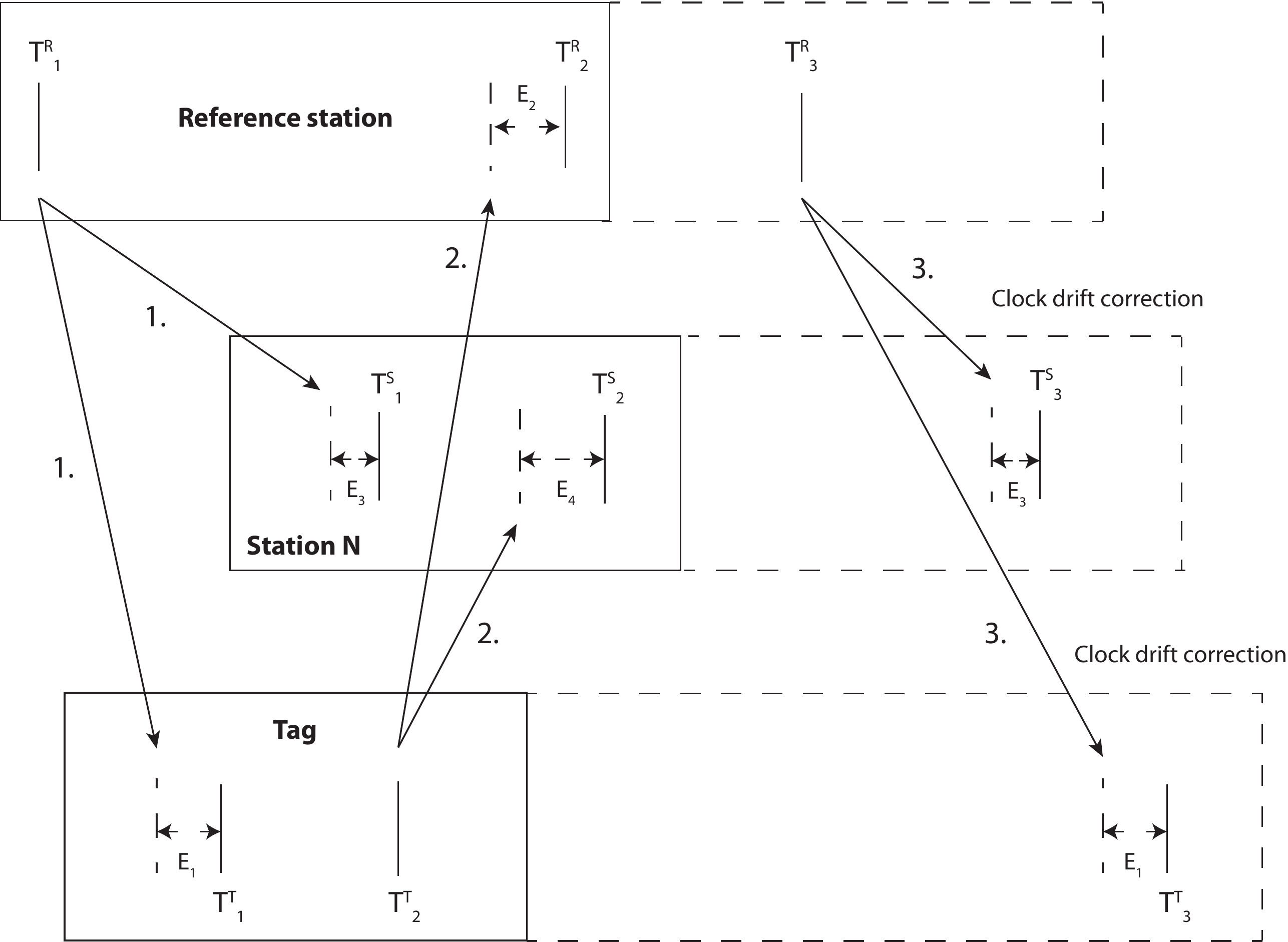}
\par\end{centering}
\caption{TOA and TDOA clock drift correction \label{fig:TDOA-1}}
\end{figure}

The previous section showed how the clock drift and the hardware offset
influence the time-of-arrival position estimate. In this section,
we show how to combine TOA with TDOA. Unlike TDOA, two-way ranging
(TWR) based on TOA does not require clock synchronization. One approach
to synchronizing the TDOA clock is to use an additional signal \cite{LPM}.
This signal is already present in the two-way ranging (TWR) approach,
so a combination of both techniques seems natural. This principle
is illustrated in figure \ref{fig:TDOA-1}. The effect of the clock
drift and the hardware delay on the TDOA can be seen in figure \ref{fig:TDOA}.
Two-way ranging is performed between the tag and the reference station.
The other stations are passive and do not respond to the reference
station or tag. The difference between timestamps two and one at each
anchor depends on the positions of the reference station and the tag
with respect to the anchor. Unlike the TWR application presented earlier,
the influence of the signal power and the hardware delay differs in
the TDOA application. 

In the TDOA application, the influence of the hardware delay is assumed
to be the same for both timestamps $T_{1}^{S}$ and $T_{2}^{S}$.
Therefore, the TDOA equation is independent of the hardware delay.
However, a new offset K appears, representing the delay between the
signal of the tag with respect to the signal of the reference station.
If both stations send the signal at exactly the same time, this offset
K is zero. 

\begin{equation}
T_{TDOA_{K}}=\Delta T_{1,2}^{S}-E_{4}+E_{3}+K
\end{equation}

The third message from the reference station $\Delta T_{1,3}^{S}=T_{3}^{S}-T_{1}^{S}$
is used to calculate the clock drift error $C_{1,3}^{S}$.

\begin{equation}
C_{1,3}^{S}=\Delta T_{1,3}^{R}-\Delta T_{1,3}^{S}
\end{equation}

After performing a linear interpolation of the clock drift error $C_{1,3}^{S}$
, the TDOA equation becomes 

\begin{equation}
T_{TDOA_{K}}=\Delta T_{1,2}^{S}+\left(\frac{C_{1,3}^{S}}{\Delta T_{1,3}^{S}}\cdotp\left(\Delta T_{1,2}^{S}+E_{3}-E_{4}\right)\right)-E_{4}+E_{3}+K
\end{equation}

This equation still depends on the offset K. However, this offset
is simply the traveling time of the signal from the reference station
to the tag plus the computation time at the tag before the signal
is emitted. It may be calculated as follows:
\begin{equation}
K=T_{TOA}+\Delta T_{1,2}^{T}+\left(\frac{C_{1,3}^{RT}}{\Delta T_{1,3}^{T}}\cdotp\left(\Delta T_{1,2}^{T}+E_{1}\right)\right)+E_{1}+2B
\end{equation}

The new TDOA equation after eliminating the offset K and including
all corrections is as follows: 

\[
T_{TDOA}=\left(\frac{C_{1,3}^{S}}{\Delta T_{1,3}^{S}}\cdotp\left(\Delta T_{1,2}^{S}+E_{3}-E_{4}\right)\right)+0.5\left(\Delta T_{1,2}^{T}+\frac{C_{1,3}^{RT}}{\Delta T_{1,3}^{T}}\cdotp\left(\Delta T_{1,2}^{T}+E_{1}\right)\right)
\]

\begin{equation}
+\Delta T_{1,2}^{S}+0.5\cdotp\Delta T_{1,2}^{R}-A+B+0.5\left(E_{1}-E_{2}\right)+E_{3}-E_{4}
\end{equation}

From each measurement, we can now obtain one TOA equation and multiple
different TDOA equations, depending on the number of anchors. This
method supports high update rates with just four stations for localization
in a two-dimensional space -- two anchors, one reference station,
and one tag. 

\[
T_{TOA_{1}}\cdotp c_{0}=\sqrt{\left(x_{R_{1}}-x_{T}\right)^{2}+\left(y_{R_{1}}-y_{T}\right)^{2}}
\]

\begin{center}
\begin{tabular}{c}
$T_{TDOA_{i}}\cdotp c_{0}=\sqrt{\left(x_{T}-x_{S_{i}}\right)^{2}+\left(y_{T}-y_{S_{i}}\right)^{2}}-\sqrt{\left(x_{R_{1}}-x_{S_{i}}\right)^{2}+\left(y_{R_{1}}-y_{S_{i}}\right)^{2}}$\tabularnewline
$1\leq i\leq N$\tabularnewline
\end{tabular}
\par\end{center}

This equation is not symmetric due to the dependency on the noise
of reference station. The reference station should therefore be selected
to have the lowest possible noise. We recommend readers to refer to
our previous publication on symmetric TDOA equations \cite{ION_Paper2}.

\section{Two-dimensional position estimation with four stations}

In this section, the theoretical concepts are verified with real measurements.
The first test scenario uses TOA measurements to estimate the unknown
position of the tag. In the second test scenario is the position of
the tag estimated by the fused measurements of TDOA and TOA.

The tests were carried out with a Decawave EVB DW1000. The Decawave
supports different message types, which are specified for the discovery
phase, ranging phase and final data transmission. Depending on the
update rate and the preamble length, each message can vary from 190
\textmu s to 3.4 ms. In our experiments, we only used 190 \textmu s
messages, also called blink messages. The general settings of the
Decawave transceivers are listed in table \ref{tab:Test-settings}. 

\begin{table}[H]
\begin{centering}
\begin{tabular}{|c|c|}
\hline 
Channel & 2\tabularnewline
\hline 
Center Frequency & 3993.6 MHz\tabularnewline
\hline 
Bandwidth & 499.2 MHz\tabularnewline
\hline 
Pulse repetition frequency & 64 MHz\tabularnewline
\hline 
Preamble length & 128\tabularnewline
\hline 
Data rate & 6.81 Mbps\tabularnewline
\hline 
\end{tabular}
\par\end{centering}
\caption{Test settings \label{tab:Test-settings}}
\end{table}

\begin{figure}[H]
\begin{centering}
\includegraphics[scale=0.4]{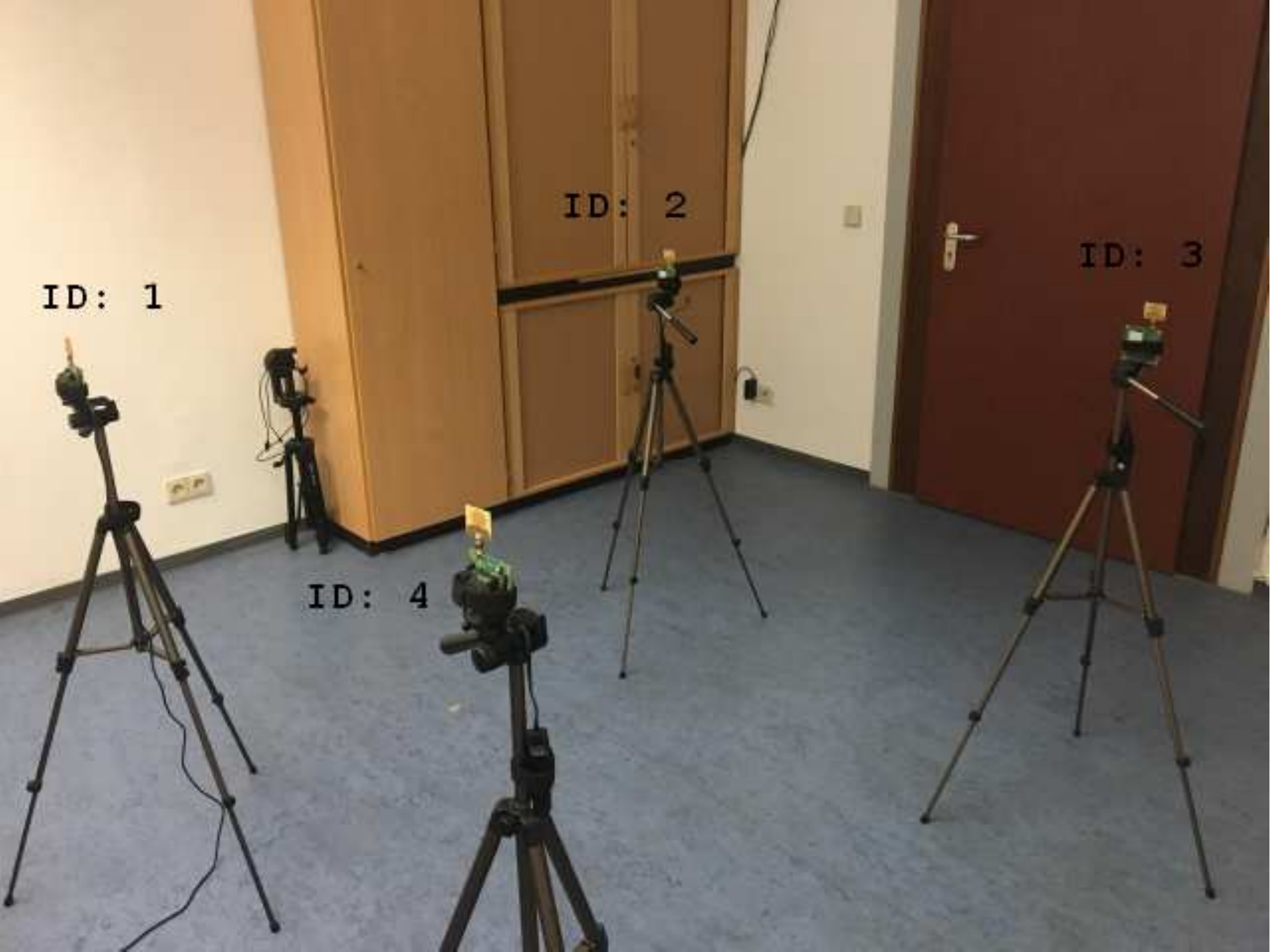}
\par\end{centering}
\caption{Constellation of the stations \label{fig:Constellation-of-the}}
\end{figure}

Figure \ref{fig:Constellation-of-the} and table \ref{tab:Position-of-the}
show the constellation of the stations. The ground truth data were
obtained by laser distance measurement. The position of the tag with
identification number (ID) 2 is assumed to be unknown. The other stations
are used to estimate the position of this tag. The station identified
as the reference station changes during TWR positioning. This is because
the distances between the tag and the other stations must be calculated
successively for TWR trilateration. Unlike TWR, the reference station
remains the same for TDOA; in this example, the reference station
is the station with ID 1. This also explains why TDOA is much faster
than TWR. 

\begin{table}[H]
\begin{centering}
\begin{tabular}{|c|c|c|}
\hline 
Station ID & X-Axis {[}m{]} & Y-Axis {[}m{]}\tabularnewline
\hline 
1 & 0 & 0\tabularnewline
\hline 
2 & 0 & 1.5134\tabularnewline
\hline 
3 & 1.27 & 1.643\tabularnewline
\hline 
4 & 1.1439 & 0.0385\tabularnewline
\hline 
\end{tabular}
\par\end{centering}
\caption{Position of the stations obtained by laser distance measurements \label{tab:Position-of-the}}
\end{table}

.

\begin{figure}[H]
\begin{centering}
\includegraphics[scale=0.6]{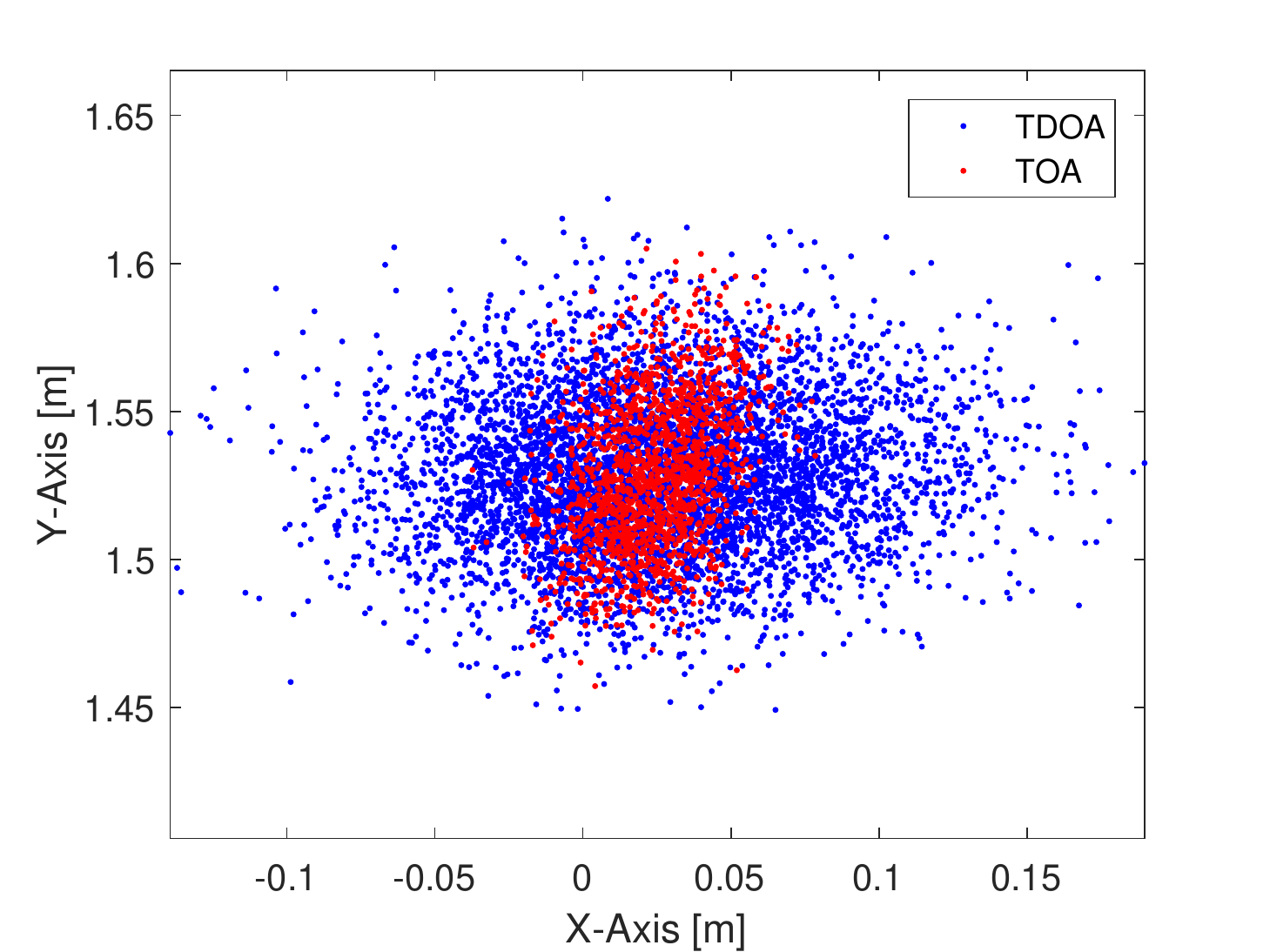}
\par\end{centering}
\caption{X/Y positions of the TOA and TDOA fused with TOA position estimates
\label{fig:Results-for-TOA}}
\end{figure}

Figure \ref{fig:Results-for-TOA} shows the results of the TOA and
TDOA position estimate of station 2. The mean values of TOA and TDOA
differ by 0.0023 m on the x-axis and 0.0006 m on the y-axis. This
difference is small, indicating that the assumptions regarding the
offset and the clock drift are correct. The deviation between the
mean values of the TOA and TDOA measurements and the ground truth
data may be explained by uncertainty in the hardware delay and the
ground truth data estimate. 

The following table \ref{tab:Precission:-Standard-Deviation} shows
the standard deviation of the precision of the TOA and TDOA position
estimates. The y-axis scattering is almost exactly equal for both
measurement techniques. On the other hand, the x-axis scattering of
TDOA is higher than that of TOA, depicted in Table \ref{tab:Covariance-matrix-of}. 

\begin{table}[H]
\begin{centering}
\begin{tabular}{ccc}
$Cov\left(TDOA\right)=\left(\begin{array}{cc}
0.0023 & 0.0001\\
0.0001 & 0.0007
\end{array}\right)$ &  & $Cov(TOA)=\left(\begin{array}{cc}
0.0003 & 0.0001\\
0.0001 & 0.0006
\end{array}\right)$\tabularnewline
\end{tabular}
\par\end{centering}
\caption{Covariance matrix of the TOA and fused TDOA measurements\label{tab:Covariance-matrix-of} }
\end{table}

This effect is due to the asymmetry of the TDOA, which is actually
a fusion of TWR and TDOA. An alternative reference station would change
the distribution. The compensation of this effect is described in
a previous publication \cite{ION_Paper2}. When combined with a filter,
highly accurate results can be obtained. The position of the anchors
affects the tag localization; better results are obtained with tags
that are more centered with respect to the anchors \cite{Placement}.

\begin{table}[H]
\begin{centering}
\begin{tabular}{|c|c|c|}
\hline 
 & TOA & TDOA\tabularnewline
\hline 
X-axis {[}m{]} & 0.0175 & 0.0479\tabularnewline
\hline 
Y-axis {[}m{]} & 0.0249 & 0.0256\tabularnewline
\hline 
\end{tabular}
\par\end{centering}
\caption{Precision: Standard Deviation \label{tab:Precission:-Standard-Deviation}}
\end{table}

The accuracy depends on the true position of the anchors and the offset
estimate. This topic will be explained in detail in an upcoming publication. 

\section{Conclusion}

This paper introduces a method of clock drift, signal power dependency,
and hardware delay correction for measurements based on the time of
arrival and the time difference of arrival. We showed how wireless
clock calibration can be performed for the time difference of arrival
using an additional station. The corrected time of arrival and time
difference of arrival measurements were combined to increase the number
of equations for the time difference of arrival position estimate.
The final section of the paper examined the theoretical concepts presented
in previous sections against real measurements from Decawave EVK1000
UWB transceivers. 

\bibliographystyle{unsrt}
\bibliography{uwb_tdoa}

\end{document}